\documentclass[pre,aps,twocolumn,superscriptaddress,amsmath,showpacs,floatfix]{revtex4}

\usepackage{graphicx}
\usepackage{amssymb} 

\begin{document}

\title{Pulling Knotted Polymers}

\author{Oded Farago}
\email{farago@mrl.ucsb.edu}
\affiliation{School for Physics and Astronomy, Tel Aviv
University, Tel Aviv 69978, Israel} 
\affiliation{Materials Research Laboratory, University of
  California, Santa Barbara, CA 93106, USA}
\author{Yacov Kantor}
\email{kantor@orion.tau.ac.il}
 \affiliation{School for Physics and Astronomy, Tel Aviv
University, Tel Aviv 69978, Israel}
 \affiliation{Department of Physics, Massachusetts
Institute of Technology, Cambridge,
Massachusetts 02139, USA} 
\author{Mehran Kardar}
\email{kardar@mit.edu}
\affiliation{Department of Physics, Massachusetts
Institute of Technology, Cambridge,
Massachusetts 02139, USA} 

\begin{abstract}
\vspace{.35cm} 

We compare Monte Carlo simulations of knotted and unknotted polymers
whose ends are connected to two parallel walls.  The force $f$ exerted
on the polymer is measured as a function of the separation $R$ between
the walls.  For unknotted polymers of several monomer numbers $N$, the
product $fN^\nu$ is a simple function of $R/N^\nu$, where $\nu\simeq
0.59$.  By contrast, knotted polymers exhibit strong finite size
effects which can be interpreted in terms of a new length scale
related to the size of the knot.  Based on this interpretation, we
conclude that the number of monomers forming the knot scales as $N^t$,
with $t=0.4\pm 0.1$.

\end{abstract}

\maketitle

Entanglements are unavoidable in long polymers and influence their
properties \cite{catenanes}.  Knots are found in proteins
\cite{taylor}, and present an obstacle that needs to be overcome in
the transcription of DNA \cite{alberts}.  An increasing number of
experiments can now probe the detailed properties of knotted molecules
\cite{dna}. Micro--manipulation techniques \cite{mechsing} enable
direct measurements of mechanical properties of a single molecule, and
it is even possible to probe the behavior of artificially knotted DNA
\cite{arai}.  However, incorporation of topological constraints into
the statistical mechanics of polymers \cite{topcon} remains a
difficult theoretical challenge since the resulting partition of phase
space into accessible and inaccessible regions cannot be easily
implemented.  Nevertheless, some progress has been made in
understanding the role of knots in loop polymers; e.g. the relative
probabilities for appearance of different knots in self--avoiding (SA)
loops \cite{knotprob} has been characterized.  However, much less is
known about the typical shapes and physical properties of knots.

\begin{figure}[ht]
\includegraphics[width=8cm]{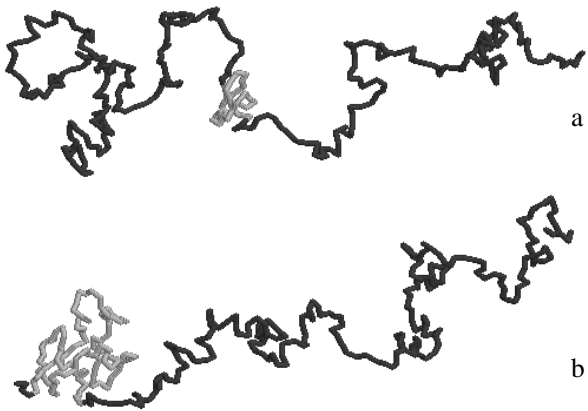}
\caption{Two configurations of a knotted chain of $N=335$ bonds,
with its ends attached to parallel walls separated by
50$a$. The walls are not shown and the chains were rotated
for clarity of view. The (grey) knotted regions consist of 32 
bonds in (a) and 112 bonds in (b).}
\label{knots}
\end{figure} 

As a topological feature, a knot can be rigorously identified only by
specifying the {\em entire} shape of a closed chain (ring).  However,
it is natural to identify a segment where the knot is located, and
consequently to talk about its size and statistics.  It is natural to
pull on the ends of a string to see if a small knot remains in the
middle.  Our eyes tend to identify knotted segments in this manner, as
exemplified by the grey--shaded monomers of the polymers in
Fig.~\ref{knots}.  While pulling on a string makes tight knots, the
question of whether in a random {\em unforced} chain the knot is
spread over the whole curve, or localized (tight) on a small portion,
is still not fully resolved.  Several recent works \cite{metzler} show
that knots in SA walks confined to a two-dimensional (2D) plane are
{\em strongly}\/ localized, i.e. the mean number of monomers in the
knotted region $N_k$ does not depend on the total number of monomers
$N$.  Studies of three--dimensional (3D) knotted polymers are hampered
by the difficulty of identifying the knotted region.  Katritch et
al.~\cite{katritch} examined the size distribution of knots in 3D
random rings, by removing segments of the ring, attaching them to
infinite straight lines, and checking if the resulting structures were
knotted.  Although the method has a certain probability to fail (i.e.,
the procedure itself may create or remove a knot), it nevertheless
suggests that knots in such rings are localized.  By contrast, some
studies \cite{largeknots} had earlier indicated that for moderate $N$,
the radius of gyration $R_g$ of a knot is strongly influenced by its
complexity, leading to the conclusion that the knots might be spread
out over the entire loop \cite{stretch}.  More recent numerical
results \cite{janse,orlandini} provide evidence that $R_g$ is
asymptotically independent of the knot type, hinting that knots are
localized.

Since correlation functions of SA walks are power laws, it is
reasonable to expect that the distribution of the number of monomers
$n$ forming a knot is also a power law, i.e. $p(n)=Cn^{t-2}$.  For
$t<1$ the normalization coefficient $C$ is determined by the
microscopic (short distance) properties of the chain, while for
$0<t<1$, the expectation value of the mean knot size $N_k$ depends on
the total length, growing as $N^t$.  In such a case, we say that the
knot is {\em weakly}\/ localized.  For $t< 0$, the knot is {\em
  strongly} localized in the sense that $N_k$ is determined by the
microscopic cutoff.  In this paper, we attempt to quantify the
tightness of knots in 3D polymers by comparing the force--extension
relations of knotted and unknotted chains.  As in
Ref.~\cite{orlandini}, we find similarities between the statistical
properties of knotted chains and those of shorter unknotted chains, a
consequence of the fact that knotted segments are statistically denser
than unknotted ones.  Using the reduction in the effective number of
monomers as an operational definition of knot size, we find $N_k\sim
N^t$, with $t=0.4\pm 0.1$.

Much of the current understanding of the scaling properties of long
polymers is based on renormalization group ideas \cite{rgpolymers}.
The $N\rightarrow\infty$ limit in polymers corresponds to approaching
a fixed point similar to that describing criticality in thermal phase
transitions.  In particular, it can be shown that for a polymer with
either free ends or forming a closed loop (without any further
restrictions on topology)
\begin{equation}
R_g=aN^\nu \Phi\left(\frac{N_0}{N},\frac{N_1}{N},\dots\right)
\simeq AaN^\nu\left[1-BN^{-\Delta}\right],
\label{rg}
\end{equation}
where $a$ is a microscopic length--scale (of the order of monomer
diameter or bond length), while $\nu\approx0.59$ (in 3D) is a system
independent (universal) exponent.  The function $\Phi$ includes
corrections to the leading power--law due to irrelevant variables,
which can be interpreted as additional length scales $\{N_i\}$.
Keeping only the largest such correction for large $N$ leads to the
second part of the above equation.  The dimensionless constants $A$
and $B$ are again system specific, while $\Delta\approx0.5$ in 3D is
universal \cite{rgpolymers}.  If the leading correction comes from,
say, the argument $N_0/N$ of $\Phi$, the corresponding length--scale
grows as $N_0\sim N^{t_0}$, with $\Delta=1-t_0$.

For very long polymers, the correction term in Eq.~(\ref{rg}) is
irrelevant, and the dependence of many physical properties on the
number of monomers $N$ can be cast in terms of a dependence on $R_g$
\cite{degennes}.  For example, consider the force $f$ needed to
stretch a polymer between two parallel walls at a distance $R$.  We
can construct two dimensionless quantities, $fR_g/k_BT$ and $R/R_g$
($T$ is the temperature and $k_B$ is the Boltzmann constant), which
must be functionally related.  It is thus convenient to introduce
variables $f'\equiv faN^\nu/k_BT$ and $R'\equiv R/aN^\nu$, and express
the force--extension relation in the form
\begin{equation}
f'=G(R').
\label{scaling}
\end{equation}
Simple arguments \cite{degennes} can now be used to determine the
asymptotic behaviors of $G(R')$: For a large stretching force, the
distance $R$ between the ends of the polymer must be proportional to
$N$. Conversely, in a strongly compressed state ($R\ll R_g$), the
force must be proportional to $N$.  These limiting behaviors can be
reconciled with Eq.~(\ref{scaling}) only if \cite{degennes}
\begin{equation}
\label{asympt}
G(R')\sim\left\{
\begin{array}{ll}
  {R'}^{\nu/(1-\nu)} & \rm{for\ } R'\gg1, \\
  -{R'}^{-1-1/\nu}   & \rm{for\ } R'\ll1.
\end{array}
\right.
\end{equation}
In the large force regime this gives $R\sim
aN(fa/k_BT)^{(1-\nu)/\nu}=aN_b^\nu(N/N_b)$ (omitting dimensionless
prefactors), with $N_b\equiv (k_BT/fa)^{1/\nu}$.  Thus, the polymer
can be viewed as a linear sequence of $N/N_b$ {\em blobs}, 
each of size $aN_b^\nu$ and 
consisting of $N_b$ monomers \cite{pincus}.  On
length--scales smaller than the blob size, the external forces are not
significant, while on the length--scales larger the polymer is
essentially linear. (An analogous blob picture is also available for
the compressed regime \cite{degennes}.)

We employed Monte Carlo (MC) simulations to measure such
force--extension relations.  Our model chains were composed of hard
spheres of diameter $0.75a$ connected by ``tethers'' restricting the
distance between adjacent spheres to be smaller than $a$, with no
additional energy costs.  The end--monomers were fixed to two infinite
parallel walls a distance $R$ apart.  An elementary MC step consisted
of an attempt to move a randomly chosen sphere a distance $0.16a$ in a
random direction.  ($N$ such attempts constitute one MC time unit.)
With such parameters the chain cannot cross itself, and its topology
is preserved by the impenetrable walls.  The force $f$ was calculated
from the probability densities of contacts between spheres, and the
probability densities of having stretched tethers, as described in
Ref.~\cite{farago}.

\begin{figure}[ht]
\includegraphics[width=8cm]{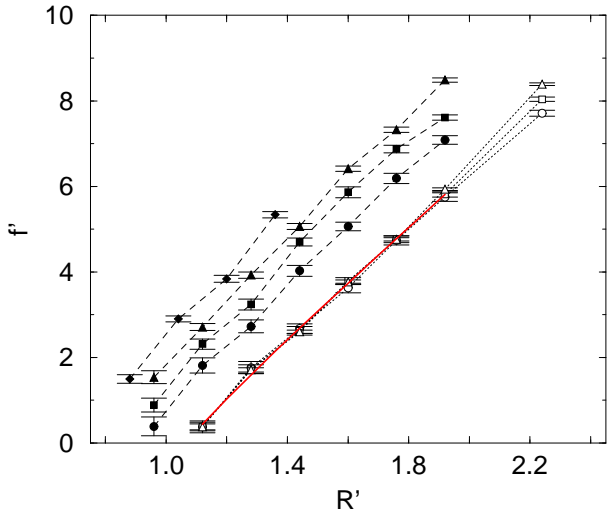}
\caption{The scaled force $f'$ as a function of the scaled separation 
 between walls $R'$, for chains of sizes $N=750$ (circles), $N=500$
 (squares), $N=335$ (triangles), and $N=225$ (diamonds) monomers. Open 
 and solid symbols correspond, respectively, to unknotted and knotted
 chains. The solid line depicts an analytical fit to $G$ in this range.}
\label{force}
\end{figure}

We studied chains of lengths $N=225$, 335, 500, and 750, in both
unknotted (simply connected), and knotted (connected via a single
trefoil) states \cite{remark}.  Fig.~\ref{knots} depicts two different
configurations of the knotted chains, where the region in which the
knot is ``concentrated'' has been lighter shaded. For each $R$ and $N$
our simulations lasted about $10^8$ MC time units, which is
considerably longer than the estimated Rouse relaxation time
\cite{doi_edwards}. We verified that during the relaxation period the
knot is able to diffuse from one side to the other.  Open and closed
symbols in Fig.~\ref{force} depict the results for unknotted and
knotted chains, respectively.

It is important to note that the usual derivation of
Eqs.~(\ref{rg})-(\ref{asympt}) is not for a specified polymer
topology, but rather for an ensemble of polymers that includes all
possible topologies. However, relations of this type are also likely
to be valid for polymers of fixed topology, albeit with a different
scaling function in Eq.~(\ref{rg}).  The collapse of the data for {\em
  unknotted} polymers (open symbols) of different lengths in
Fig.~\ref{force} confirms this expectation.  Indeed, the quality of
the collapse indicates that subleading corrections are negligble for
$R'<2$. We have taken advantage of this observation to construct an
analytic fit to the scaling function $G$, as depicted by the solid
line in this figure.  For $R'>2$ the polymer with $N=335$ forms more
than six blobs, each containing less than 55 monomers. For such small
blobs, finite size effects begin to appear, as we observe a roughly
10\% deviation in the right--most data points in Fig.~\ref{force}.
Note that a deviation of this magnitude is expected from Eq.~\ref{rg},
with $B\sim 1$.

As expected, for a given $f'$ the knotted polymer (solid symbols in
Fig.~\ref{force}) has a significantly smaller $R'$ than its unknotted
counterpart.  However, the scaled difference becomes less pronounced
for larger $N$, and the results for longer chains approach those of
the unknotted chain.  The absence of data collapse indicates the
appearance of strong finite size corrections in a parameter range
where the unknotted chains show no such effect.  We would like to
associate this feature with the emergence of a new size scale due to
the knot.  If a string with a knot is fully stretched, its maximal
length is reduced by the size of the resulting tight knot.  While the
knots in our simulations are far from tight, we shall still describe
the influence of the knot as a reduction in the number of monomers $N$
by the `size of the knot' $N_k$.  If the knotted chain of length $N$
is equivalent to an unknotted chain of length $N-N_k$, its
force--extension curves must satisfy $f a(N-N_k)^\nu/k_B T
=G[R/a(N-N_k)^\nu]$, where $G$ is the scaling function obtained before
for unknotted polymers (solid line in Fig.~\ref{force}).  Naturally,
this definition will not work with a single $N_k$, since our knots are
not tight.  However, each solid data point in Fig.~\ref{force} can be
moved to the previously obtained solid line by choosing an appropriate
$N_k(f',N)$.  This can be regarded as our operational definition of
the size of the knot for a given $f$ and $N$.

If the force--extension curves for knotted polymers are to be
consistent with the standard finite size corrections discussed in
connection with Eq.~(\ref{rg}), we must have
\begin{equation}
f'=G_k\left(R',\frac{N_0}{N},\dots\right)\simeq
G(R')\left[1+g(R')N^{-\Delta_k}\right].
\label{fwithcor}
\end{equation}
Based on the numerical results, we have assumed that the leading
scaling function $G$ is the same for knotted and unknotted polymers,
but allowed for different corrections, such as a new exponent
$\Delta_k$.  For Eq.~(\ref{fwithcor}) to be consistent with our
definition of knot size, we must have $N_k(f',N)=H(f')N^t$.  The
exponent $t=1-\Delta_k$ gives the scaling of the number of monomers in
a knot (in the absence of force) via $N_k\sim N^t$.  General scaling
considerations do not restrict the shape of $H(f')$.  However, from
the ``blob picture'' of a strongly stretched chain, we know that the
polymer is essentially linear beyond the blob size $N_b$ (and
consequently not knotted on such scale), while within a blob it is
undisturbed by the external forces. We thus expect the knot size to be
determined by $N_b$ (as if this is the entire length of the polymer),
i.e. $N_k\sim N_b^t=N^t/{f'}^{t/\nu}$, and $H(f')\sim 1/{f'}^{t/\nu}$
for $f'\gg1$.  Similar behavior occurs \cite{mkk} in force--extension
characteristics of polymers in which a sliding constriction (or
slip-link \cite{sliplink}) creates a loop, somewhat reminiscent of a
topological constraint. In the latter, the size of the loop under
strong tension is equal to the size of a loop in unstressed polymers
of size $N_b$.

\begin{figure}[ht]
\includegraphics[width=8cm]{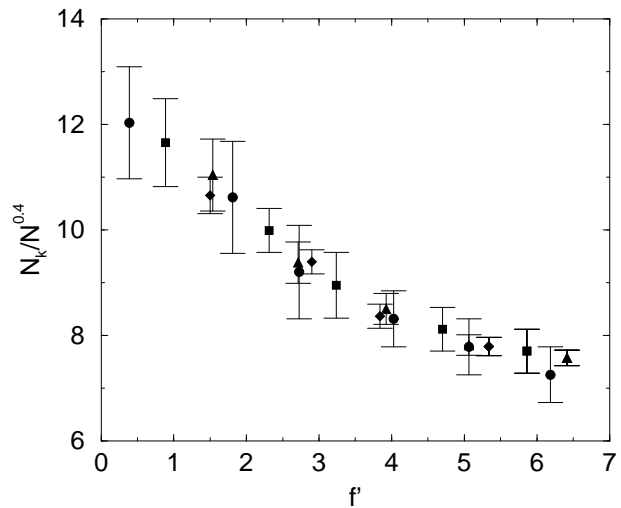}
\caption{The scaled knot size $N_k/N^{0.4}$, as a function of the 
  scaled force $f'$, for chain lengths $N=750$ (circles), 500
  (squares), 335 (triangles), 225 (diamonds).}
\label{scale}
\end{figure}  

While our range of extensions is too limited to test the asymptotic
behavior of $H(f')$, we can estimate the exponent $t$ from the value
at which the functions $N_k/N^t$ exhibit the best collapse, i.e.~are
least sensitive to $N$.  Fig.~\ref{scale} depicts the optimal collapse
with $t=0.4$, which is characterized by a $\chi^2$ value of $0.35$.
For $t=0.3$ and $t=0.5$ we have $\chi^2\simeq 1$, serving as a
criterion for the error in $t$, and our estimate of $t=0.4\pm 0.1$.
This is somewhat smaller than the value that can be deduced from
Ref.~\cite{orlandini}.  The corresponding finite--size correction
exponent is $\Delta_k=0.6$.  We note that this is close to the best
numerical estimate $\Delta\simeq 0.56$, for the dominant correction to
scaling when considering all topologies \cite{li}, and within errors,
also is consistent with the estimate $\Delta\simeq 0.48$ obtained by
field theoretic techniques\cite{guida}.  Is this more than simple
coincidence?  The standard field theory for polymers is based on an
expansion around four dimensions that does not incorporate topological
constraints.  Assuming that the analytic continuation of this theory
to three dimension also tells us about knots, how can such effects be
anticipated in the perturbative expansion?  If knots do indeed appear
with subleading sizes in 3D, their effects could be anticipated in
corrections to scaling, in which case $\Delta_k=\Delta$.  This
conjecture is bolstered by $\Delta\simeq\Delta_k\simeq 1$ for knots
confined to 2D \cite{metzler,li}.

In summary, by comparing force--extension relations of unknotted and
knotted polymers of several lengths, we observe strong finite-size
corrections in the latter, which we attribute to the knot size $N_k$.
Scaling analysis and data collapse suggest a power law $N_k\sim N^t$,
with $t=0.4\pm 0.1$.  Thus unlike 2D `flat knots' \cite{metzler}, the
3D knot sizes grow with the length of the polymer (although as a
diminishing fraction of the whole length).  Since a single knot is
only weakly localized and ``knows'' about the size of the chain, it is
interesting to investigate chains with several knots which may
interact with each other.  This, however, requires simulations with
much larger chains.

We thank C.~Jeppesen for helpful comments.  Simulations were performed
at High Performance Computing Unit of the Inter University Computation
Center of Israel. This work was supported by the National Science
Foundation (DMR-01-18213 and PHY99-07949) and the US-Israel Binational
Science Foundation (1999-007). OF acknowledges the support of the MRL
Program of the National Science Foundation (DMR00-80034).
%----------------------------------------------------------- 
%References
%-----------------------------------------------------------


\begin{thebibliography}{99}
\bibitem{farago} O. Farago and Y. Kantor, Phys. Rev. E {\bf 61}, 2478
  (2000). To measure the force we use Eq.(14) in this paper with the
  prefactor $1/V$ replaced by $1/R$.
\bibitem{remark} Knots are traditionally defined on closed rings. To
  obtain an open chain with a trefoil one has to cut the ring at one
  of the arcs between two crossing points.
\end{thebibliography}
\end{document}